\documentclass[12pt]{iopart}

\expandafter\let\csname equation*\endcsname\relax

\expandafter\let\csname endequation*\endcsname\relax

\usepackage{amsmath,amssymb,calrsfs}
\usepackage{graphicx}
\usepackage{subfigure}
\usepackage[colorlinks=true,citecolor=blue,urlcolor=blue,linkcolor = blue]{hyperref}  
\usepackage{dcolumn}
\usepackage{bm}
\usepackage{verbatim}
\usepackage{units}
\usepackage[english]{babel}
\usepackage[utf8]{inputenc}
\usepackage{upgreek}
\usepackage{comment}

\usepackage[dvipsnames]{xcolor}

\usepackage[normalem]{ulem}

\newcommand{\beq}{\begin{equation}}
\newcommand{\eeq}{\end{equation}}
\newcommand{\beqa}{\begin{eqnarray}}
\newcommand{\eeqa}{\end{eqnarray}}

\begin{document}

\title[Soliton trains after interaction quenches in Bose mixtures]{Soliton trains after interaction quenches in Bose mixtures}

\author{Andr\'e Cidrim}
\address{Departamento de F\'isica, Universidade Federal de S\~ao Carlos, 13565-905 S\~ao Carlos, Brazil.}

\author{Luca Salasnich}
\address{Dipartimento di Fisica e Astronomia ``Galileo Galilei", Universit\`a di Padova, via Marzolo 8, 35131 Padova, Italy}
\address{INFN, Sezione di Padova, via Marzolo 8, 35131 Padova, Italy}
\address{CNR-INO, via Nello Carrara, 1 - 50019 Sesto Fiorentino, Italy}
\address{Padua Quantum Technologies Research Center, University of Padova, Via Gradenigo 6/b, 35131 Padova, Italy}

\author{Tommaso Macr\`i}
\address{Departamento de F\'isica Te\'orica e Experimental, Universidade Federal do Rio Grande do Norte, and International Institute of Physics, Natal-RN, Brazil}

\begin{abstract}
We investigate the quench dynamics of a two-component Bose mixture and study the onset of
modulational instability, which leads the system far from equilibrium. Analogous to the single-component counterpart, this phenomenon results in the creation of trains of bright solitons.
We provide an analytical estimate of the number of solitons at long times after the quench for each of the two components based on the most unstable mode of the Bogoliubov spectrum, which agrees well with our simulations for quenches to the weak attractive regime when the two components possess equal intraspecies interactions and loss rates. We also explain the significantly different soliton dynamics in a realistic experimental homonuclear potassium mixture in terms of different intraspecies interaction and loss rates.
We investigate the quench dynamics of the particle number of each component estimating the characteristic time for the appearance of modulational instability for a variety of interaction strengths and loss rates. Finally we evaluate the influence of the beyond-mean-field contribution, which is crucial for the ground-state properties of the mixture, 
in the quench dynamics for both the evolution of the particle number and
the radial width of the mixture. In particular, even for quenches to strongly attractive effective interactions we do not observe the dynamical formation of solitonic droplets.
\end{abstract}


\maketitle

\section{Introduction}
\label{sec:introduction}

Modulational instability (MI) is a generic phenomenon that consists of the spontaneous exponential growth of perturbations resulting from the interplay between nonlinearity and anomalous dispersion.
MI occurs in several areas of physics. It has been observed in classical~\cite{benjamin_feir_1967,doi:10.1146/annurev.fl.12.010180.001511} and quantum \cite{Strecker:2002aa,Nguyen2017,Everitt2017,PhysRevA.99.033625,PhysRevA.66.063602} fluids, in waveguides \cite{PhysRevLett.56.135} and in lattices \cite{PhysRevLett.90.140405}, as well as in nonlinear optics \cite{agrawal2019} and in charged plasmas \cite{THORNHILL197843}.

In ultracold Bose-Einstein condensates several experimental \cite{Strecker:2002aa,Nguyen2017,Everitt2017,PhysRevA.99.033625,PhysRevA.66.063602} and
theoretical \cite{Salasnich_PhysRevLett.91.080405,PhysRevLett.89.170402,PhysRevA.70.033607,PhysRevLett.93.100401,PhysRevE.87.032905} works examined the conditions for the appearance of MI. 
In such systems the combination of dissipative nonequilibrium 
dynamics and the nonlinearity due the interactions results in MI that, after a variable time interval, induces the formation of a train of solitons \cite{Strecker:2002aa,Nguyen2017,Everitt2017,PhysRevLett.89.200404,  Kiehn2019,Mistakidis2018}. 
On the theoretical side, the far-from-equilibrium dynamics induced by an interaction quench from the repulsive regime to the attractive is usually well captured by mean-field 
approaches based on the solution of the Gross-Pitaevskii equation with the inclusion of dissipative three-body losses.
When the BEC is confined in a quasi one-dimensional waveguide a description based on the non polynomial nonlinear Schrödinger equation typically predicts accurately the number of solitons for relatively weak
attractive interactions \cite{Salasnich_PhysRevA.66.043603,Everitt2017}.
The relevant scale that determines the insurgence of MI is the (inverse) most unstable wavenumber $q_{MI}$. 
Then, one finds that the number of solitons increases monotonically with the final value of the final scattering length.
Interestingly, it is also possible to observe that the particle
number of the BEC decreases with a universal power law of the holding time
rescaled by the characteristic time for the creation of the modulational
instability, independently of the strength of the quench \cite{Nguyen2017}. Recently, the excitation spectrum of matter-wave solitons has also been measured in a quasi one-dimensional cesium condensate \cite{Di_Carli_2019}.

Whereas most works on MI in BECs focused on a single component BEC, the realization of 
multicomponent systems offers a natural playground to observe nonequilibrium effects in a more general framework.
Restricting to two-component BECs, one notices already a 
rich variety of phases in the ground state. 
In purely repulsive mixtures,
one observes a homogeneous superfluid or a phase separation when 
inter-species repulsion overcomes the intra-species interaction strength.
In the attractive regime a series of recent experiments showed the formation of dilute self-bound droplet states in a two-component BEC both in a tight optical waveguide \cite{Cabrera301,Cheiney2018} and in free space \cite{Semeghini2018,PhysRevLett.106.065302,PhysRevResearch.1.033155}, closely following the theoretical predictions \cite{Petrov2015}.
In the quasi one-dimensional geometry, upon varying the mean-field interaction from the weakly to the strongly attractive regime, one observes a smooth crossover between bright soliton states and self-bound droplets. 
Notably, solitons are excitations appearing genuinely in low-dimensional systems. If the one-dimensional interaction strength is attractive (focusing nonlinearity) one retrieves bright solitons, while for repulsive condensates (self-defocusing nonlinearity) one finds dark solitons.
Droplets instead result from the competition between mean-field and quantum fluctuation energies with opposite sign.

Although the ground-state properties of binary BEC mixtures have been extensively investigated, the study of the conditions leading to MI in these setups has only recently received attention, both theoretically \cite{Malomed2020sym12010174} and experimentally \cite{Cheiney2018}.
MI has been observed in the counterflow dynamics of two-component BECs in the miscible (purely repulsive) phase \cite{PhysRevLett.106.065302}.
Also, a recent experiment with coherently coupled BECs rapidly quenched into the attractive regime reported the creation of bright soliton trains formed by dressed-state atoms \cite{sanz2019interaction}.

In this work we thoroughly investigate the quench dynamics in a binary mixture of Bose-Einstein condensates from the repulsive to the attractive regime in an elongated quasi one-dimensional waveguide. We provide results for quenches from the repulsive to the weakly and strongly attractive regime, where solitonic states and quantum droplets respectively are expected in the ground state. We quantitatively characterize the resulting nonequilibrium dynamics by computing the number of particles and solitons as a function of the holding time after the quench.
In section~\ref{sec:model} we present a theoretical model for a two-component mixture which allows us to provide a quantitative estimate of the number of solitons following a quench dynamics based on the most unstable mode.
In section~\ref{sec:results} we discuss our numerical results.
We perform simulations
for the quench dynamics in three different regimes: repulsive to soliton, repulsive to droplet and soliton to droplet. Finally in section~\ref{sec:conclusions} we present our conclusions. 
In the appendix we provide some details about the ground-state phase diagram, the dynamics of the radial width of the two-component mixture, the effect of beyond-mean-field corrections on the nonequilibrium dynamics and the algorithm used to monitor the number of solitons in our simulations.

\section{Model and simulations}
\label{sec:model}
We describe the nonequilibrium dissipative dynamics of a binary homonuclear mixture in a quasi one-dimensional waveguide with radial trapping frequency $\omega_r$ and components $i,j=1,2$ (with $i\neq j$) by the generalized coupled Gross-Pitaevskii equations (GPEs), which, in rescaled units, read

\begin{equation} \label{eGPE}
  \begin{aligned}[t]
    i\frac{\partial\psi_i}{\partial t}  = & \left(-\frac{1}{2}\nabla^2+V(\mathbf{r})+4\pi \frac{a_i}{a_r} \left|\psi_i\right|^2 + 4\pi \frac{a_{ij}}{a_r}\left|\psi_j\right|^2 +\right.\\
    & \left. \frac{128\sqrt{\pi}}{3}\frac{a_{i}}{a_r}\left(\frac{a_{i}}{a_r}\left|\psi_i\right|^2+
    \frac{a_{j}}{a_r}\left|\psi_j\right|^2\right)^\frac{3}{2} -i\Gamma_i\left|\psi_i\right|^4 \right)\psi_i.
    \end{aligned}
\end{equation}

\noindent Here time, energy and space are scaled in units of
the inverse radial trapping frequency $\omega_r^{-1}$, the radial trapping energy $\hbar\omega_r$ and the associated harmonic oscillator length $a_r=\sqrt{\hbar/m\, \omega_r}$, respectively and sum over the indices in the interaction term is assumed. 
The coefficients $a_i$ and $a_{ij}$ furnish the intra- and inter-species corresponding scattering lengths. The effective mean-field scattering length is thus given by the parameter $\delta a = a_{12}+\sqrt{a_{1}a_{2}}$. The number of particles in each component equals $N_{1,2}$. We consider a cigar-shaped trapping harmonic potential $V(\mathbf{r})=m\left[\omega_r^2(x^2+y^2)+\omega_z^2 z^2\right]/2$ with frequencies $\omega_r/2\pi= 346\,\mathrm{Hz}$ and $\omega_z/2\pi = 7.6\,\mathrm{Hz}$, similarly to \cite{Nguyen2017}, thus having a radial harmonic oscillator length of $a_r=0.87\, \mu$m. 
The generalization comes first phenomenologically including a three-body loss term $\Gamma_i=L^{(i)}_3 N_i^2/(2\omega_r a_r^6)$. Following \cite{Semeghini2018} we do not include mixed two-body or three-body inelastic loss rates.
Secondly, in order to take into account corrections due to quantum fluctuations, we introduce the Lee-Huang-Yang (LHY) terms in the GPEs, with component-dependent strength.
The ratio between the mean-field chemical potential and the transverse harmonic oscillator energy $(g_1 n_1+g_2 n_2)/\hbar \omega_r\sim 1$ for typical parameters used in this work, which allows us to use consistently the three-dimensional LHY expression \cite{Gajda2018_PhysRevA.98.051603}. Note that for purely one-dimensional dynamics an attractive quantum fluctuation term would appear \cite{Malomed2020sym12010174}.

We perform simulations in two different regimes: 
(i) symmetric and (ii) realistic. 
In (i) we set equal initial intraspecies scattering lengths $a_{1i}=a_{2i}=53.0\,a_0$ and variable final interactions with the constraint $a_{1f}=a_{2f}$, where subscripts $i$ and $f$ indicate values before and after quench. We also set $a_{12} = a_{12i} = a_{12f} = -52.0\,a_0$
and fix the three-body losses and particle numbers to the same value for each component to $L_3=3.6\times 10^{-27}\mathrm{cm^6/s}$.
Case (ii) is motivated by current experiments on dilute quantum droplets, with homonuclear mixtures made of two hyperfine states of $^{39}K$, $\left|F=1,m_f=0\right>$ and $\left|F=1,m_f=-1\right>$~ \cite{Semeghini2018,Cheiney2018}. We consider the interval of magnetic fields $56\, \text{G}<B<57\, \text{G}$, where for the two hyperfine states the intraspecies scattering lengths $a_{1}$ and $a_{2}$ are both positive (repulsive). 
The interspecies scattering length $a_{12}$, instead, is negative. In the setup we are studying the initial intra-component scattering lengths are chosen to be $a_{1i} = 75.0\,a_0$ and $a_2 = a_{2i} = a_{2f} = 37.5\,a_0$. Component 1 presents a largely varying scattering length $a_1$ for a particular experimentally tunable range (via Feshbach resonances), while $a_2$ and the inter-species scattering length $a_{12} = a_{12i} = a_{12f} = -52.0\,a_0$ remain practically constant for the same range \cite{Cabrera301,Cheiney2018,Semeghini2018}. The three-body loss rates $L_3^{(1)}=3.6\times 10^{-27}\mathrm{cm^6/s}$ and $L_3^{(2)}=6.0\times 10^{-29}\mathrm{cm^6/s}$~ \cite{Semeghini2018}.
\begin{figure}
	\centering
	\includegraphics[width=0.65\columnwidth]{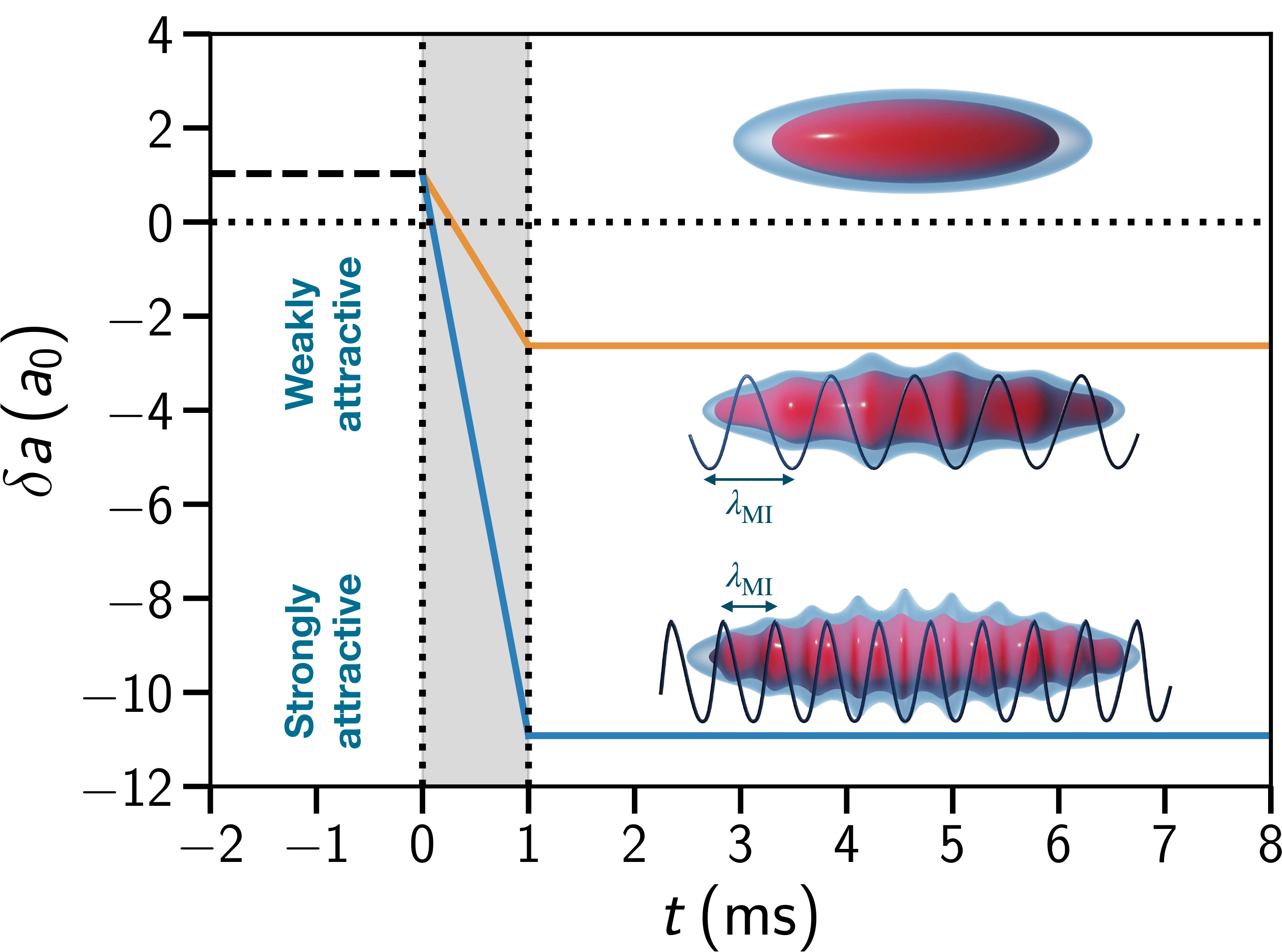}
    \caption{Quench protocol for a two-component BEC mixture from the repulsive to the attractive regime. 
	At time $t=0$ the effective scattering length $\delta a$ is quenched using a linear ramp of the duration of $\delta t = 1\, \text{ms}$ from $\delta a = \delta a_{12}+\sqrt{a_{11}a_{22}}>0$ to the attractive regime with $\delta a<0$ either in the soliton phase or to the strongly attractive in the droplet phase. Within the same time interval we switch off the longitudinal confinement. This allows the system to expand freely and to monitor the creation of bright soliton trains propagating along the $z$-direction. 
	We also consider the quench from the solitonic state to the self-bound droplet which is discussed in the text.}
	\label{Fig:droplet}
\end{figure}

\begin{figure}[!ht]
    \centering
    \includegraphics[width=0.7\columnwidth]{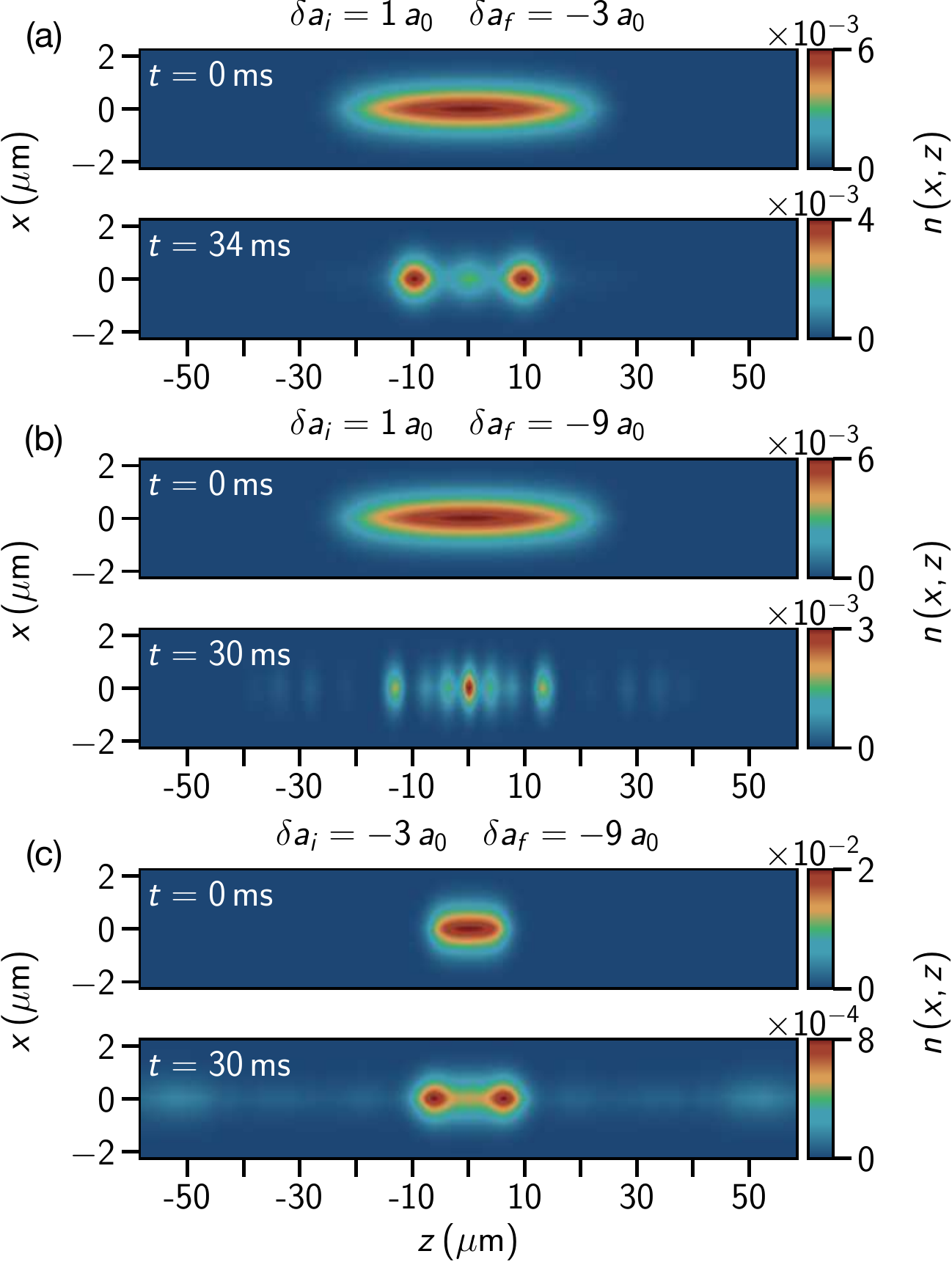}
    \caption{Snapshots of the time-evolution of the density at the initial time $t=0$ and at $t=30$ ms after a quench with a $1$ms-ramp for different initial and final scattering lengths. We consider the initial number of particles $N_1=N_2=2.5\times 10^4$ and symmetric intraspecies interactions $a_1=a_2$ which are quenched simultaneously. The interspecies scattering length $a_{12}$ is left unaltered during the quench. The system is initialized in the ground state for the corresponding value of $\delta a_i$ by relaxing in imaginary time a variational gaussian wavefunction. For details of the numerical algorithm and the initial ground-state function see \ref{GS_diagram}.
    The three-body loss parameters are chosen equal to $L_3^{(1)}$.
    After the quench the initial state develops a MI towards the soliton train visible at longer times.
    (a) Quench from the repulsive ground state at $\delta a_i=1\, a_0$ to the soliton regime at $\delta a_f=-3\,  a_0$.
    (b) Quench to the droplet phase at $\delta a_f=-9\, a_0$.
    (c) Quench from the soliton $\delta a_i=-3\, a_0$ to the droplet phase at $\delta a_i=-9\, a_0$. See supplementary material (\href{https://stacks.iop.org/NJP/23/023022/mmedia}{https://stacks.iop.org/NJP/23/023022/mmedia}) for videos showing the full dynamics of
the three cases presented here.}
    \label{fig:time_ev_panel}
\end{figure}

In the absence of a harmonic confining potential and neglecting beyond-mean-field effects one would obtain a dilute gas phase for $\delta a>0$ and a collapsing BEC for $\delta a<0$. The addition of the beyond-mean-field contribution to the equation of state of the two-component system leads to the stabilization of self-bound quantum droplets due to the competition of attractive mean-field terms proportional to $n^2$ and the repulsion due to LHY-type terms proportional to $n^{5/2}$ in the corresponding equation of state \cite{Cappellaro2018}.
The introduction of the harmonic trap leads to a rich new physical phenomenology. The detailed ground-state phase diagram for the parameters studied in this work is described in~\ref{GS_diagram} together with the details of the numerical algorithm to obtain the ground states. The algorithm to investigate the soliton dynamics is presented in \ref{sec:algorithm}.

In most cases, when choosing initial states for our simulations, we start from repulsive inter-species interactions (i.e. $\delta a>0$) before quenching to attractive values. The quench is performed by a rapid variation of the scattering length with a linear ramp of $1$ms. Concurrently, we switch off the longitudinal trapping potential and observe the system expanding in a time-of-flight fashion. See figure~\ref{Fig:droplet} for a schematic representation of the quench protocol. We also notice that the small aspect ratio of our trapping potential (i.e. $\omega_z/\omega_r \ll 1$) implies that the most relevant dynamics will happen along the axial direction. In our simulations of the GPE in equation~(\ref{eGPE}) we assume cylindrical symmetry. This choice reduces the computation to an effective two-dimensional calculation of $\psi(r,z)$, over a grid size of $16\times 8192$ points and a domain of $5\,a_r\times 1290\,a_r$ allowing to resolve in detail the dynamics along the axial direction. 
In order to evaluate the derivatives involved in the kinetic term, we employ a discrete (zero-order) Hankel transform in the radial direction and the usual fast Fourier transform in the longitudinal direction \cite{footnote}.

\section{Results}
\label{sec:results}

\subsection{Modulational instability in a binary Bose mixture}
\label{mod_instability}
In this section we present a theoretical model to describe the MI in a binary BEC after a quench to the attractive regime $\delta a_f<0$.
To characterize quantitatively the MI we introduce the Bogoliubov spectrum for a two-component BEC, which reads \cite{LARSEN196389}
\beq
E_\pm(q) =
\sqrt{
\frac{\varepsilon_1^2+\varepsilon_2^2}{2}\pm
\sqrt{\frac{(\varepsilon_1^2-\varepsilon_2^2)^2}{4}+
\frac{g_{12}^2 n_1 n_2 q^4}{m_1 m_2}}
},
\eeq
where we defined the Bogoliubov energies of each component 
\beq 
\varepsilon_i(q) = 
\sqrt{\frac{q^2}{2m_i} \left(\frac{q^2}{2m_i}+g_{ii}n_i\right)}.
\eeq
In this work we focus on the equal mass case $m_1=m_2$. We define the total density of the system as $n=n_1+n_2$. 
In our simulations we fix the ratio of the densities (and particle numbers) 
at the initial time to
$\frac{n_1}{n_2} = \sqrt{\frac{a_{2}}{a_{1}}}$, i.e. the equilibrium configuration that minimizes the energy functional in the attractive regime
\cite{PhysRevResearch.2.013269}.
Under the condition $\delta a = a_{12}+\sqrt{a_{1}a_{2}}<0$
the lower branch $E_-(q)$ becomes unstable. The most unstable mode corresponds to the wavenumber $q_\text{min}$ that minimizes the argument of $E_-(q)$.

Upon solving $d E_-(q)/dq=0$ we obtain
\beqa \label{nsolitons}
q_\text{min}^2
&=& 2 n \sqrt{a_{1}a_{2}}\left(
\sqrt{1- \frac{8\,\delta a}{(\sqrt{a_{1}}+\sqrt{a_{2}})^2}+
\frac{4\,\delta a^2}{\sqrt{a_{1}a_{2}}(\sqrt{a_{1}}+\sqrt{a_{2}})^2}}
-1
\right),
\eeqa
\noindent where we scaled the scattering lengths $a_i$ and consequently also $\delta a$ by a factor $2\pi$ to account for the effective soliton dynamics along the longitudinal direction.
When $|\delta a|\ll(\sqrt{a_{1}}+\sqrt{a_{2}})^2$ then
we can expand $q_\text{min}$ to a reduced expression to first order in $\delta a$ which reads
\beqa \label{nsolitonsred}
q_\text{min}^2\approx(q_\text{min}^\text{red})^2 
&=&  8 n\, |\delta a| 
\frac{\sqrt{a_{1} a_{2}}}{(\sqrt{a_{1}}+\sqrt{a_{2}})^2}.
\eeqa

We can now define the associated wavelength $\lambda_0=2\pi/q_\text{min}$. 
Starting with a quasi-1D binary Bose mixture of length $L$ 
(ex. the longitudinal Thomas-Fermi radius), 
a sudden quench to the attractive regime 
induces the formation of a train of solitons. 
The number of solitons $N_s$ can then be estimated as
\beq\label{ns_lambda0}
N_s = \frac{L}{\lambda_0}. 
\eeq 

\subsection{Number of solitons}
\begin{figure}[!t]
	\centering
	\includegraphics[width=0.6\columnwidth]{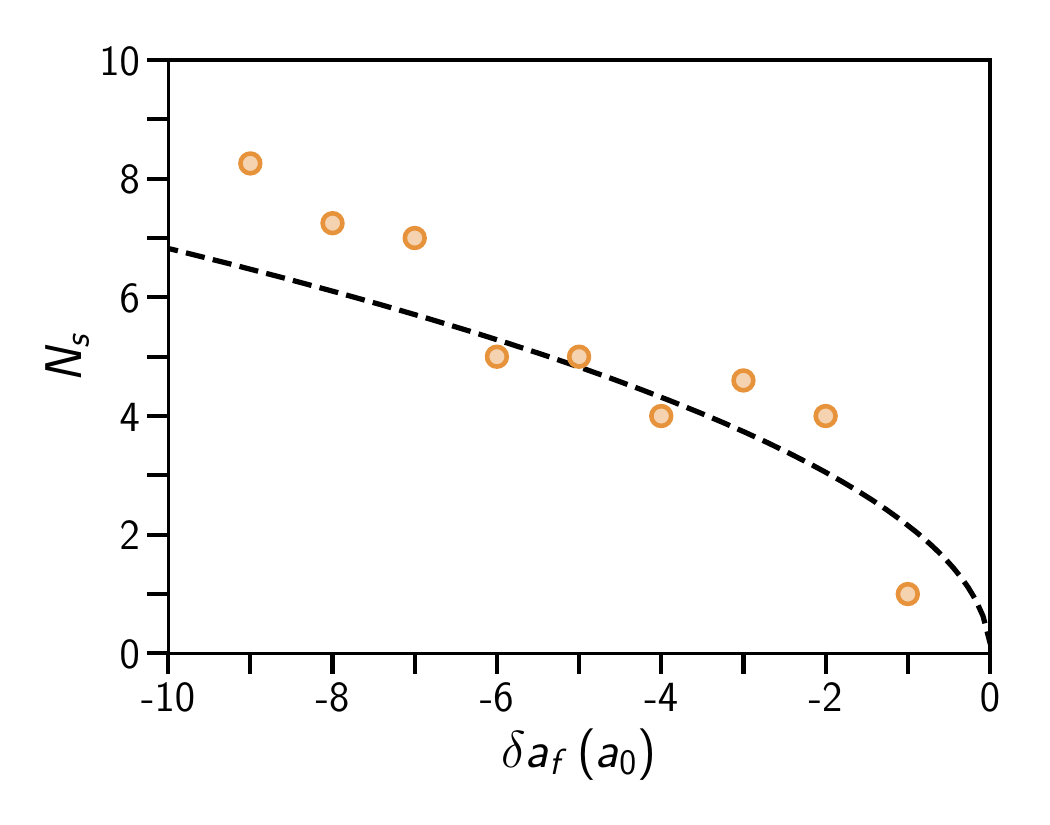}
	\caption{Number of bright solitons in quench dynamics as a function of the final value of the effective scattering length $\delta a_f$ for a symmetric mixture with equal intraspecies interactions and dissipations. Solitons are identified from counting density peaks after radial integration. 
	We compare the numerical results with the estimate of the number of solitons from equations~(\ref{nsolitons}) and (\ref{ns_lambda0}) (black dashed line).
	We set $N_1=N_2=2.5\times 10^4$ particles with $\omega_r/2\pi = 346$ Hz and $\omega_z/2\pi = 7.6$ Hz and $L_3^{(1)}=L_3^{(2)}=3.6\times 10^{-27}\mathrm{cm^6/s}$. Here we set $a_{12}=-52\, a_0$ and vary $a_{1}=a_{2}$, computing then $\delta a = \delta a_{12}+\sqrt{a_{1}a_{2}}$. The dynamics is initialized in the ground state with a repulsive BEC with $a_{1}=a_{2}= 53\, a_0$.}\label{fig:Ns_t_and_delta_af}
\end{figure}

We now discuss the creation of soliton trains induced by the MI. In figure~\ref{fig:time_ev_panel} we show snapshots of the density for the symmetric regime (i) at two different times: one right before the quench from the repulsive mean-field regime to the attractive one and another at $t=30$ms. Figure~\ref{fig:time_ev_panel}(a) corresponds to a quench to the weakly attractive regime $\delta a=-3\, a_0$ where the ground state of the system is an extended soliton (see~\ref{GS_diagram} for a thorough discussion of the phase diagram). Figure~\ref{fig:time_ev_panel}(b) corresponds to a quench to the strongly attractive regime $\delta a=-9\, a_0$ where the ground state is instead a self-bound droplet. For both cases the initial state is the ground state of the system for 
repulsive interactions. 
Notice that for the sake of visualization the density 
is normalized at its peak value, therefore the color code is
the same for both components to emphasize the density modulations, even if the
number of particles changes with time. We observe that, as soon as MI sets in, density peaks are formed, creating a soliton train. The number of solitons increases with the strength of the attractive interaction.

We compute the number of solitons numerically from the peaks of the density distribution for each of the two components, employing an algorithm that we describe in detail in \ref{sec:algorithm}. 
The results of this analysis are shown in figure~\ref{fig:Ns_t_and_delta_af}.

In figure~\ref{fig:Ns_t_and_delta_af} the average number of solitons computed from the numerics (points) is compared to our prediction from equations (\ref{nsolitons}) and  (\ref{ns_lambda0}) (dashed black line).
\noindent We notice that the expression for $q_\text{min}^{\mathrm{red}}$ in equation~(\ref{nsolitonsred}) reproduces to an excellent approximation $q_\text{min}$ for the parameters used in this work.
For weakly attractive interactions the agreement between the numerics and the analytical result is good for $|\delta a|\lesssim 6\, a_0$. 
For stronger attractive interactions we observe a larger deviation of the analytical prediction from the numerics, likely due to the far from equilibrium dynamics involved in the creation of the soliton train.

In figure~\ref{fig:time_ev_panel}(c) we show the snapshot of the density starting from a solitonic, weakly attractive configuration to the regime of strong attraction. We observe that the initial condensate splits 
into just two bright solitons. This has to be compared to  
figure~\ref{fig:time_ev_panel}(b), where, due to the larger initial longitudinal length, the quench dynamics 
produces a soliton train with several density peaks.

We also performed simulations in the realistic case (ii) for 
the experimental parameters of section~\ref{sec:model}. 
The dynamics is significantly more complex than in the symmetric case.
First the asymmetry in the number of particles 
(see section~\ref{sec:num-part} and inset of 
figure~\ref{fig:realistic_L3_N_t}) is such that the particles in 
the second component is almost constant during the expansion dynamics
after the quench, whereas $N_1(t)$ is greatly reduced after tens of milliseconds, similarly to the symmetric case. The effect is that the second component is only weakly affected by the attractive dynamics due to the limited overlap with the first component. Therefore the soliton trains observed after the quench are poorly described by the theory described in section~\ref{sec:results}. We provide further details in \ref{sec:algorithm}.

\subsection{Number of atom loss}
\label{sec:num-part}
\begin{figure}[!h]
	\centering
	\includegraphics[width=\columnwidth]{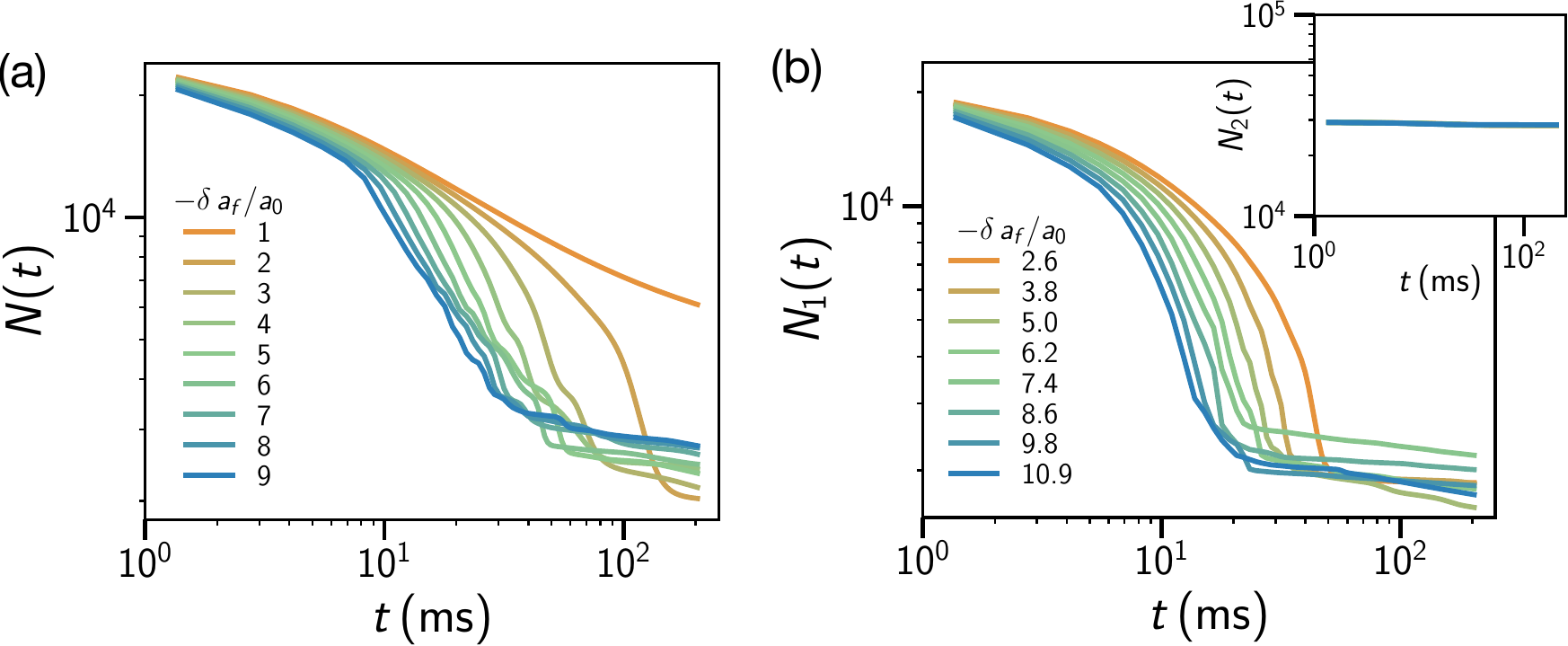}
	\caption{Time evolution for number of atoms, starting from a total $N=5.0\times 10^4$ particles for the case (a) of symmetric intraspecies interactions $a_1=a_2$ and equal three-body loss coefficients $L_3^{(1)}=L_3^{(2)}=3.6\times 10^{-27}\mathrm{cm^6/s}$ and (b) realistic case of asymmetric intraspecies interactions and dissipations as described in section~\ref{sec:model}. 
	The system is quenched from $\delta a_i = 1\,a_0$ to different $\delta a_f$. Notice that for realistic three-body losses coefficients $L_3^{(1)} \gg L_3^{(2)}$, therefore the second component (inset) practically maintains its particle number throughout the evolution.
	In (a) the dissipation acts on the two components equally, therefore $N_1(t)=N_2(t)$ for all times. Different curves in (a) and (b) correspond to different final effective interaction strength as in the color scale bar.
    }
	\label{fig:realistic_L3_N_t}
\end{figure}

In this section we discuss the evolution of the number of particles as a function of time after the quench. 
In figure~\ref{fig:realistic_L3_N_t} we show the results for both quantities from GPE simulations of the two coupled components $N_1(t)$ and $N_2(t)$ for (a) the case of symmetric interactions and losses (b) and for the realistic experimental parameters of section~\ref{sec:model}.
In the symmetric case $N_1(t)=N_2(t)$ for all times, whereas in the realistic case the number of particles in components $1$ and $2$ decrease with time at different rates because of the different three-body losses coefficients $L_3^{(1)},L_3^{(2)}$.
The first component has a much larger three-body decay, resulting in a more complex dynamics.
We observe that for all the final attractive mean-field interactions considered in figure~\ref{fig:realistic_L3_N_t}, the number of particles for short times slowly decreases before establishing the MI at $t \approx t_{MI}$
\beqa \label{t_MI}
t_{MI}\, \omega_r &=&
\frac{(\sqrt{a_{1}}+\sqrt{a_{2}})^2}{\sqrt{a_{1}a_{2}}}
\frac{1}{4 n_0 |\delta a| a_r^2}.
\eeqa
where $n_0$ is the peak density of the initial 
configuration
\cite{Salasnich_PhysRevA.66.043603}.
Consistently with the prediction of equation (\ref{t_MI}), in our simulations for the symmetric (a) and the realistic case (b),
quenching to the strongly attractive regime leads to faster decrease of the number of particles.

\section{Conclusions}
\label{sec:conclusions}
In this work we studied the nonequilibrium dynamics of a two-component Bose-Einstein condensate after a quantum quench to the attractive interspecies interactions. We specialized to the experimentally relevant case of potassium binary mixture. Quenching the effective mean-field scattering length from repulsive to attractive values in a wide interval we observed a MI and the creation of soliton trains. We characterized quantitatively the number of solitons via numerical simulations of the coupled Gross-Pitaevskii equations. In the stationary, long-time limit we observed that an analytical model based on the calculation of the most unstable Bogoliubov mode is in reasonable quantitative agreement with the number of solitons for both components in the symmetric configuration. The experimentally relevant case, with asymmetric intraspecies interactions and different loss rates, leads to a more intricate dynamics which is only qualitatively captured by our model. The related time scale for the rise of the instability however does not translate into a universal scaling for the losses of both components, in contrast to what was recently observed for a single component lithium BEC with small final scattering length $|a_f|$ \cite{Nguyen2017}.

We emphasize that this work focuses on a far-from-equilibrium dynamical regime. For the fast magnetic field ramps considered here, even in the strongly attractive regime $|\delta a_f|\gtrsim 7\, a_0$, the solitonic bumps in the density are not self-bound droplets, as their width equals the transverse harmonic oscillator length and the atom number decay is different from what has been observed in the formation of self-bound droplets
\cite{PhysRevResearch.2.013269} (see also~\ref{radial_width}).

The MI analysis can be used to study also other Bose-condensed systems. A sudden quench of the s-wave scattering length can be applied not only to atomic gases in the same or different hyperfine states, but also to heteronuclear bosonic mixtures. Moreover, in bosonic systems with spin-orbit and Rabi couplings the MI can be induced by varying these one-body couplings. However, our work strongly suggests that generically one cannot trust only the analytical calculations based on the most unstable mode of the elementary excitations: a comparison with numerical simulation is needed to obtain reliable predictions. Extensions of this work may include a systematic study of the effects of the coherent coupling of a two-components mixture \cite{Cappellaro2017, sanz2019interaction,PhysRevA.92.063606}, the inclusion of long-range dipolar interactions \cite{PhysRevA.97.011604}, or the investigation of finite-temperature effects \cite{Lee_2018} across the normal-to-BEC transition in the attractive regime and its connection to the Kibble-Zurek mechanism \cite{PhysRevA.100.033618,PhysRevLett.122.040406}.

\ack

We thank D. Luo, R. Hulet, and S. W\"uster for useful discussions. This research was developed with the help of XMDS2 software~\cite{Dennis2013}. We thank the High Performance Computing Center (NPAD) at UFRN for providing computational resources. T.M. acknowledges CNPq for support through Bolsa de produtividade em Pesquisa n.311079/2015-6. This work was supported by the Serrapilheira Institute (grant number Serra-1812-27802), CAPES-NUFFIC project number 88887.156521/2017-00. T.M. thanks the Physics Department of the University of L'Aquila for the hospitality where part of the work was done. A.C. is supported by FAPESP through Grant No. 2017/09390-7. LS acknowledges the BIRD project ``Superfluid properties of Fermi gases in optical potentials'' of the University of Padova for financial support.

\appendix

\section{Ground-state phase diagram} \label{GS_diagram}
\begin{figure}[!h]
\includegraphics[width=\textwidth]{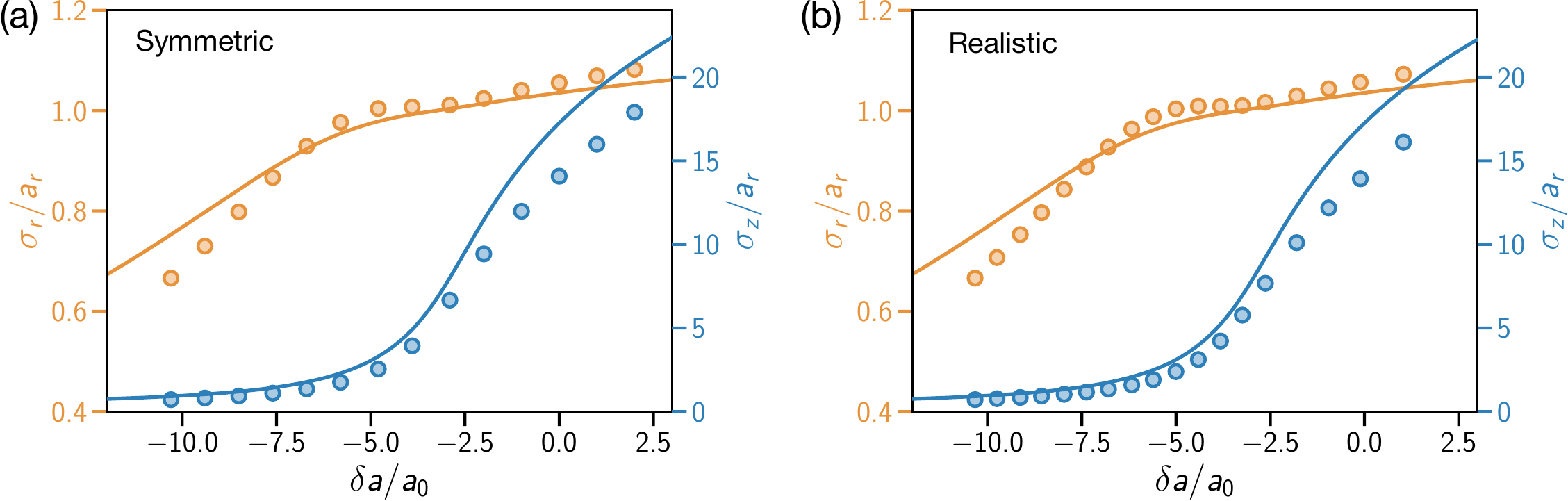}
\caption{Crossover soliton-droplet as a function of the effective scattering length for a two-component mixture with $N=5.0\times 10^4$ particles with $\omega_r/2\pi = 346$ Hz and $\omega_z/2\pi = 7.6$ Hz. Orange  line: $\sigma_r$, blue line: $\sigma_z$ in units of $a_r$. (a) Symmetric case. We set $a_{12}=-52\, a_0$ and vary $a_{1}=a_{2}$, so as to effectively vary $\delta a = \delta a_{12}+\sqrt{a_{1}a_{2}}$. (b) Realistic experimental case. We fix $a_{1}= 37.5\, a_0$ and $a_{12}=-52.0\, a_0$ and vary $a_{2}$. For large negative $\delta a$ the system is in the droplet phase $\sigma_r \sim \sigma_z \ll a_r$. For small negative $\delta a$ the system is in the soliton phase $\sigma_r \sim a_r < \sigma_z$. Numerical simulation of the ground state wavefunction obtained by imaginary-time evolution of the generalized GPE of equation~(\ref{eGPE}) in the absence of dissipation are shown with orange ($\sigma_r$) and blue ($\sigma_z$) points.
}
	\label{Fig:groundstate}
\end{figure}
In this appendix we discuss the ground-state properties of a two-component mixture in the attractive regime in a cigar-shaped harmonic potential.
In figure~\ref{Fig:groundstate} we show the numerical and the variational phase diagram obtained by imaginary-time evolution of the generalized GPE of equation~(\ref{eGPE}) (numerical) and by minimizing the corresponding energy 
functional (variational). 
For the variational approach we employ a gaussian wavefunction
\beq
\psi(\textbf{r}) = \sqrt{\frac{N}{\pi^\frac{3}{2}\sigma_r^2\sigma_z}}
\exp 
\left(-\sum_{r_i=x,y,z} 
\frac{r_i^2}{2\sigma_{r_i}^2}
\right)
\eeq
with variational parameters $\sigma_r$ and $\sigma_z$. The wavefunction is normalized to the total number of particles $||\psi||^2= N$ \cite{Cappellaro2018}.

We observe a smooth crossover from the droplet to the soliton phase. First, for low particle number or, equivalently, small values of $|\delta a|$, the ground state of the system corresponds to a soliton, whose shape depends on the external trapping, for which $\sigma_r \sim a_r$, while $\sigma_z \gg a_r$. 
Specifically, beyond-mean-field corrections are not necessary for the stability of this state \cite{Salasnich_PhysRevA.66.043603}.
Reducing $\delta a$ the ground state is (almost) isotropic $\sigma_r \sim \sigma_z < a_r$, independent of the confinement aspect ratios. 
The existence of this self-bound state is enabled by taking into account the contribution of Gaussian quantum fluctuations in the variational energy equation~(\ref{eGPE}).

We compare the ground states obtained from the variational analysis with the numerical simulation of the GP equation. After relaxation in imaginary time, we find that the ground state of the system for the strongly attractive inter-species scenario is a self-trapped, spherical droplet state.

\section{Evolution of the radial width} 
\label{radial_width}
\begin{figure}[!h]
	\centering
	\includegraphics[width=0.65\textwidth]{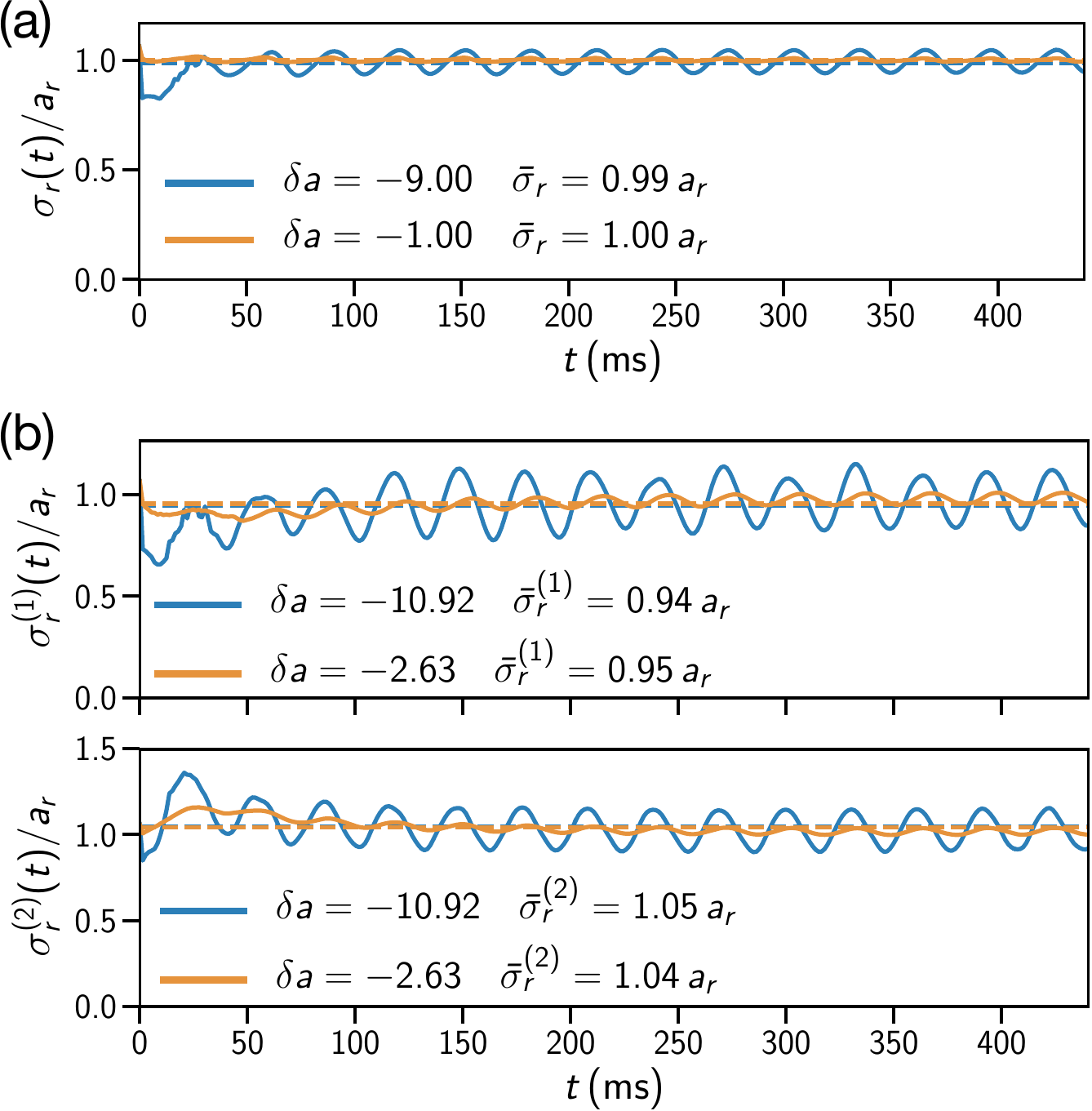}
	\caption{Time evolution of the radial widths after integrating along the longitudinal direction for the (a) symmetric and (b) realistic experimental case. In the realistic experimental case (b) we plot the time evolution for component $1$ (upper panel) and component $2$ (lower panel).
	We consider a two-component mixture with $N=5.0\times 10^4$ particles with $\omega_r/2\pi = 346$ Hz.
	We notice that, independently of the quench parameters, the radial widths oscillate around the harmonic potential length ($a_r$).}
	\label{Fig:realistic_L3_sigma_r_t}
\end{figure}

The results of the dynamics of the radial widths of each of the
two components of the mixture after the quench are shown in figure~\ref{Fig:realistic_L3_sigma_r_t}. 
The radii of both components fluctuate closely to the initial radial
harmonic oscillator length $a_r$ for both quenches to the weakly attractive and strongly attractive regimes.
In particular, for the cases where $\delta a_f= -9.0\,a_0$ (symmetric) $\delta a_f= -10.92\,a_0$ (realistic) and which have a self-bound droplet phase as a ground state (see figure~\ref{Fig:groundstate}) we observe no signature of such a state throughout the dynamics.

\section{Effect of quantum fluctuations} \label{effect_LHY}
\begin{figure}[!h]
	\centering
	\includegraphics[width=\textwidth]{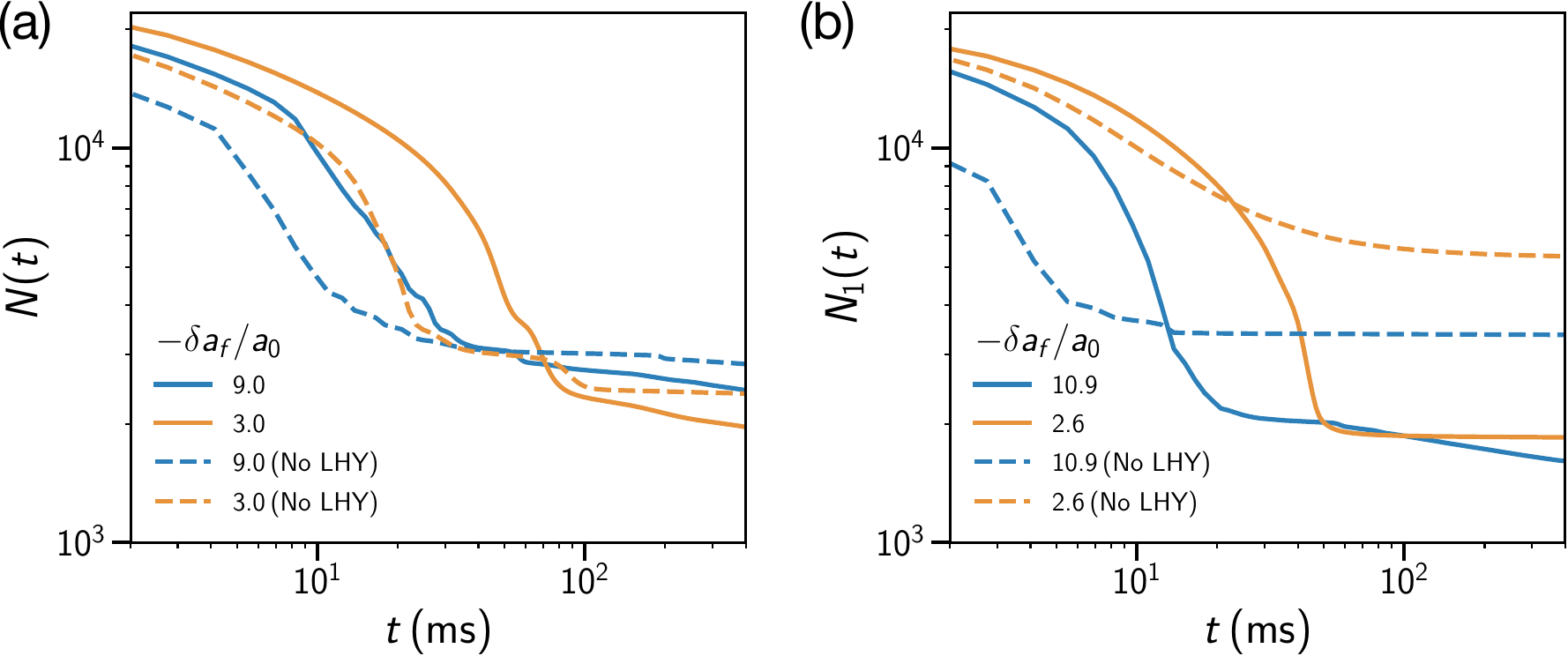}
	\caption{Effect of beyond-mean-field corrections on the dynamics of the particle number decay after the quench for (a) the symmetric case with equal interactions $a_1=a_2$ and loss rates $L_3^{(1)}=L_3^{(1)}$, and (b) the realistic experimental case with the parameters of section~\ref{sec:model}. 
	Full lines: simulation with LHY correction. Dashed lines: simulations without LHY correction. In (b) we only show $N_1(t)$ as $N_2$ remains constant for the whole time interval.}
	\label{Fig:N_t_LHY}
\end{figure}
The effect of the quantum fluctuations for two-component
Bose gas is described by the term
$g_i^{\mathrm{LHY}}=\frac{128\sqrt{\pi}}{3}\frac{a_{i}}{a_r}\left(\frac{a_{i}}{a_r}\left|\psi_i\right|^2+
    \frac{a_{j}}{a_r}\left|\psi_j\right|^2\right)^\frac{3}{2}$
in the extended GPE equation~(\ref{eGPE}).
We tested the effect of the removal of this term in our simulations
for quenches into the weak and strong attractive regime. The results are
shown in figure~\ref{Fig:N_t_LHY} for symmetric interactions 
and loss terms and for the realistic experimental configuration. 
Whereas the results for the second component in the realistic case are almost identical, the first component displays a much faster particle number decay in the absence of beyond-mean-field terms. This can be explained by the attractive nature of the mean-field interaction which is 
not balanced by the additional repulsive LHY correction leading to higher densities and therefore to higher losses. At the same time, MI takes place on a shorter time scale, leading to a faster stabilization of the particle number. Therefore, at longer times, in the absence of beyond-mean-field effects we observe a larger particle number.

\section{Details on the soliton number algorithm and comparison with theory}\label{sec:algorithm}
We briefly describe the algorithm to compute the number of solitons 
from the dynamics of the density $n(z,t)$ integrated along the transverse directions as a function of time. Our method identifies local maxima in $n(z,t)$ as bright solitons. We start by disregarding, at each instant, peaks in low-density regions (with amplitudes $\lesssim 10\%$ of the mean initial density). A weak Gaussian filter is applied to smooth out most numerical effects at distances much smaller than the typical healing length $\xi$ of the system. Finally, we avoid overcounting solitons which are undergoing a probable splitting process by treating as one visible peaks that are apart by distances smaller than $\xi$ at a particular instant.

\begin{figure}[!h]
	\centering
	\includegraphics[width=\textwidth]{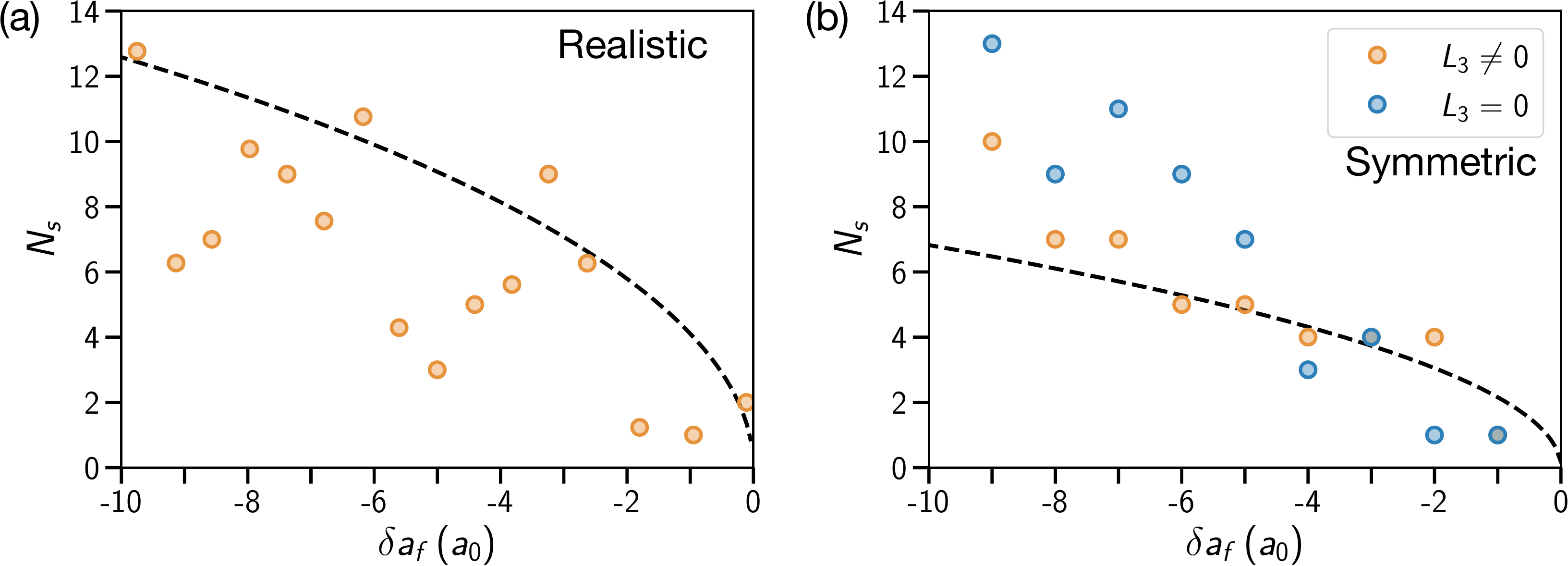}
    \caption{Number of solitons for $N=5\times 10^4$ in the (a) realistic and (b) symmetric cases. The latter compares the effect of disregarding the three-body losses.}
	\label{Fig:comparison}
\end{figure}

In figure \ref{Fig:comparison} we provide further numerical results about the comparison between the realistic and the symmetric case with our theoretical model. In the realistic case, as discussed in the main text, we only plot the number of solitons in the first component $N_s^{1}$. The dynamics in the realistic case is significantly more complex than in the symmetric one. Therefore the soliton trains observed after the quench are inadequately described by the theory in this context. Due to this poor agreement between the numerics and the estimate from equation (\ref{ns_lambda0}), we only show the effect of three-body losses on the solitons number in the more controllable symmetric case. We observe that in the absence of losses equation (\ref{ns_lambda0}) is still able to provide a reasonable estimate of the number of solitons for quenches to intermediate values of $|\delta a_f|$. However, for larger $|\delta a_f|$ the deviation can be as much as $100\%$. In the realistic case it can be argued that a similar increase in the number of solitons might be observed in the absence of dissipation, therefore partially compensating for the mismatch of the theoretical (dashed line) and the numerics. However a more refined theory is needed to explain the behavior of both components.

\section*{References}
\bibliographystyle{unsrt}
\bibliography{bose.bib,bib.bib}

\end{document}